\documentclass[twocolumn,pra]{revtex4-1}
\usepackage[T1]{fontenc}
\usepackage[utf8]{inputenc}
\usepackage{amsmath,amssymb,bbold}
\usepackage{color}
\usepackage{graphicx}
\usepackage{amssymb}
\begin{document}
\title{Magnetic Phases in Periodically  Rippled Graphene.}
\author{ M. Pilar L\'opez-Sancho and Luis Brey }
\email{Electronic address: brey@icmm.csic.es}
\affiliation{Departamento de Teor\'{\i}a y Simulaci\'on de Materiales, Instituto de Ciencia de Materiales de Madrid, CSIC, 28049 Cantoblanco, Spain}

\date{\today}

\begin{abstract}
We study the  effects that ripples induce  on the electrical and magnetic properties of graphene. 
The variation of the interatomic distance created by the ripples translates in a  modulation of the hopping parameter between carbon atoms.
A tight binding Hamiltonian including a Hubbard interaction term  is solved self consistently for ripples with  different amplitudes and periods.  
We find that, for values of the Hubbard interaction $U$ above a critical value $U_C$, the system displays a superposition of local ferromagnetic and antiferromagnetic ordered states.
Nonetheless the global ferromagnetic order parameter is zero.
The $U_C$ depends only on the product of the period and hopping amplitude modulation.
When the Hubbard interaction is close to  the critical value of the antiferromagnetic transition in pristine graphene, the  antiferromagnetic order parameter becomes much larger than the
ferromagnetic one, being the ground state similar to that of flat graphene.  
\end{abstract}

\maketitle
\section{Introduction}

Graphene is a two-dimensional material with many possibilities in  technological applications\cite{Avouris_2007,Novoselov_2012}, but also 
it presents 
many exotic and unexpected physical peculiarities\cite{Guinea_2009,Katsnelson-book}. One of the more remarkable new physical properties  of graphene  is the strain-induced pseudo magnetic gauge fields\cite{Morozov_2006,Morpurgo_2006}. In graphene uniform strain, apart from a renormalization of the Dirac velocity,  generates a constant gauge vector potential that shift the position in reciprocal space of the Dirac cones and can be gauged away. On the contrary non-uniform strain generates position dependent vector potential that induces pseudo-magnetic fields  which can be  experimentally tested. 
Non-uniform  strain occurs in  the intrinsic ripples that appear  in free standing graphene\cite{Meyer_2007,Guinea_2008,Meng_2013,Bai_2014} and also when the graphene sheet is  bonded to a substrate\cite{Stolyarova_2007,Ishigami_2007,VazquezParga_2008,Pereira_2009,Levy_2010,Lu_2012}.  Effective magnetic fields constant  on large areas can be obtained by strain engineering\cite{Guinea_2010,Guinea_2010a,Low_2010,Ramezani_2013,Roy_2014,Verbiest_2015}. The effective fields generated by non uniform strain can interfere with externally applied magnetic fields giving rise to new physical effects\cite{Prada_2010,Roy_2013,Li_2015}. Also the existence of pseudo magnetic fields could produce anomalous effects in the electronic  quantum transport\cite{Settnes_2016}.

The origin of the appearance of the gauge fields is the lineal dispersion of the bands near the Dirac points\cite{Amorim_2016} and  gauge fields have been predicted to occur in a variety of physical
systems with linear dispersion as topological insulators\cite{Tang_2014,Venderbos_2016}, optical lattices\cite{Tian_2015}, modulated graphene superlattice\cite{Sun_2010}, molecular graphene\cite{Gomes_2012} and other two-dimensional semimetals\cite{Zabolotskiy_2016}.

Previous work has shown that the application of an uniform  uniaxial strain to graphene  reduces the value of the critical  coupling constant for exchange instability towards a ferromagnetic (FM) phase\cite{Sharma_2013,Peres_2005}. Also,  the critical value of the on-site Hubbard coupling  for an  antiferromagnetic (AFM) instability was found to be reduced with respect the case of pristine graphene\cite{Viana-Gomes_2009,Sorella_1992,Martelo_1996}.   In the case of non uniform strain 
the appearance of pseudo magnetic fields induces a peak in the density of states at the Fermi energy and the instability of the system against magnetic ordering.
In this work we study, by solving self-consistently the Hubbard model,  the electrical and magnetic properties of  rippled graphene. 
 We model the graphene ripple, a lattice deformation, by a sinusoidal modulation of the hopping parameter of period $L$ and amplitude $\delta _{t}$.

\begin{figure}[htbp]
\includegraphics[width=9.5cm,clip]{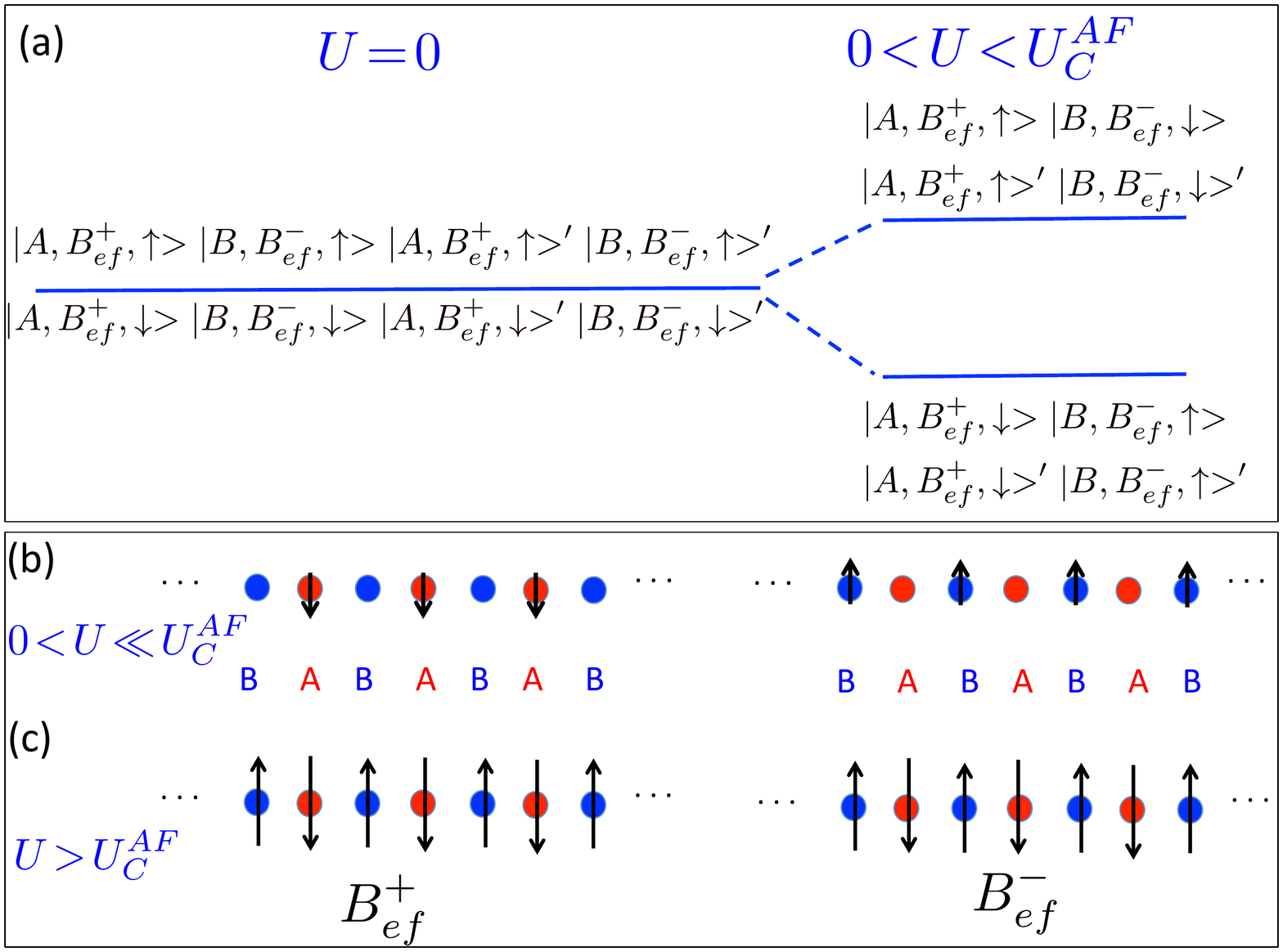}
\caption{(Color online) (a) Pseudo Landau levels appearing in rippled graphene. In absence of interactions, $U=0$, there are eight zero energy degenerated pseudo Landau levels corresponding to states located in regions with positive, $B_{ef} ^+$, or negative, $B_{ef} ^-$ effective magnetic fields,  with spin projection $\uparrow$ or $\downarrow$, and in Dirac points ${\bf K}$ or ${\bf K}'$. In regions with $B_{ef} ^{+(-)}$  the wave functions have only amplitude in sublattice $A(B)$. A moderate  interaction $U$ opens an exchange energy gap, favoring a local  FM order in  regions with $B_{ef}^+$ and opposite polarized FM order
in  regions with $B_{ef}^-$. In (b) we show schematically the real space magnetic order for moderate $U$. For larger values of the Hubbard interaction (c), rippled graphene gets a N\'eel order with a very weak FM modulation.
}                                   
\label{order}
\end{figure}

\begin{figure}[htbp]
\includegraphics[width=8.5cm,clip]{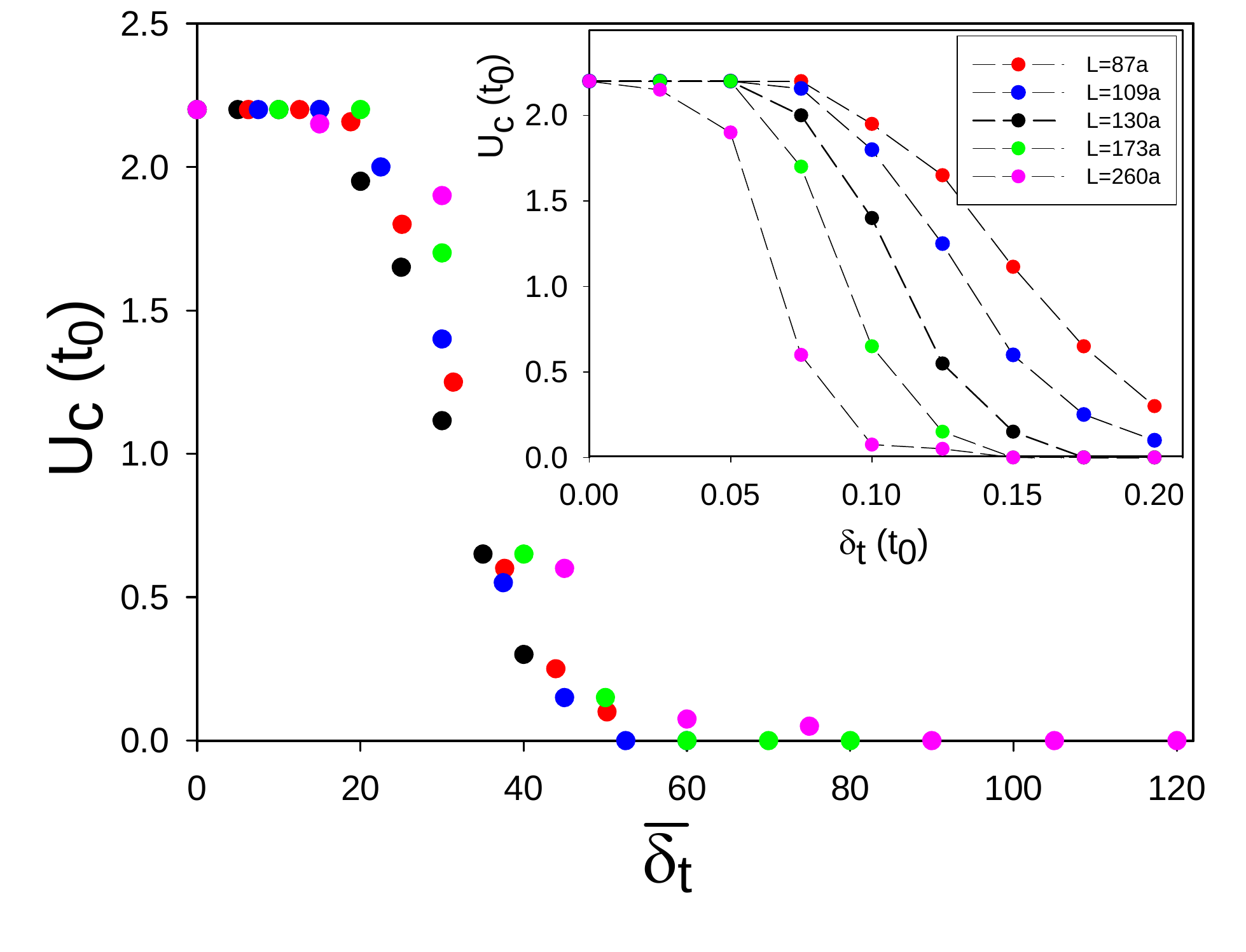}
\caption{(Color online) Critical Hubbard interaction for magnetic instabilities in rippled graphene, as function of the dimensionless parameter  $\bar \delta _t$ =$\frac {\delta _t}{\hbar v_F G}$.  Dots with different colors correspond
to different values of the period $L$. In the inset we plot $U_C$ as function of $\delta _t$ and for different periods $L$.
}                                  
\label{Phase-Diagram}
\end{figure}

The main results of our work are  summarized in Fig.\ref{order} and  Fig.\ref{Phase-Diagram}. 
In rippled graphene, for non-interacting electrons,   eight degenerated  pseudo Landau levels appear
at the Fermi energy, see Fig.\ref{order}(a),
from which only half will be occupied. For values of the Hubbard interaction greater than a critical $U_C$, the high density of states at the Fermi energy makes the system magnetically unstable and the minimization of the exchange energy opens an energy gap and selects the four occupied  pseudo Landau levels.
We obtain numerically that $U_C$ depends only on the product of the ripple amplitude by its period $\delta_t  L$, as shown in Fig.\ref{Phase-Diagram}. For small values of this parameter $U_C$ coincides practically with the critical value of the paramagnetic antiferromagnetic transition in pristine graphene, $U_C^{AF}$. 
For larger values of the product $\delta _t L$ the value of $U_C$ decreases until it reaches zero. 

For values of $U$ slightly higher than $U_C$   the system presents a 
local ferromagnetic order that correlates its polarization  with the orientation of the pseudo magnetic field, Fig.\ref{order}(b), in such a way that the total magnetization is zero.
On top of the local FM order there is an antiferromagnetic order with a small order parameter.
For larger values of $U$ the  pseudo Landau levels are destroyed and the antiferromagnetic order parameter increases
being much larger than the local FM order parameters, as schematically shown in  Fig.\ref{order}(c).

%We find that for values of the Hubbard interaction greater than a critical $U_C$ the system undergoes a phase transition  towards a magnetic ground state.

%In this  ground  state  the total magnetization is zero, 

%We obtain numerically that $U_C$ depends only on the product $\delta_t  L$. For small values of this parameter $U_C$ coincides practically with the critical value of the paramagnetic antiferromagnetic transition in pristine graphene, $U_C^{AF}$. For larger values of the product $\delta _t L$ the value of $U_C$ decreases until it reaches zero. 

%In rippled graphene there are eight degenerated  pseudo Landau levels at the Fermi energy, see Fig.\ref{order}(a).  For values of $U$ slightly higher than $U_C$  these Landau levels split
%opening an exchange gap at the Fermi energy and making the system a 
%local ferromagnetic order that correlates its polarization  with the orientation if the pseudo magnetic field, Fig.\ref{order}(b). On top of the local FM order there is a week  antiferromagnetic  N'eel order. 
%For larger values of $U$ the energy degenerated pseudo Landau levels are destroyed and the antiferromagnetic order parameter increasing behind much larger the the local FM order parameters.

Our results are consistent with the obtained in reference \cite{Roy_2014},
%The magnetic order we find is consistent with that obtained in reference \cite{Roy_2014}, 
where  a strained graphene flake is studied with an almost 
uniform axial pseudo magnetic field in the bulk that is compensated by an opposite oriented pseudo magnetic field at the edge.
In this geometry, on top of a global AFM order,   the bulk and the edge of the flake have an effective but opposite oriented magnetization\cite{Roy_2014}.

The paper is organized as follows. 
In the next section we introduce the ripple geometry and the tight-binding and Dirac like Hamiltonians describing the electronic properties of the system. Also we  refresh the concept of gauge magnetic field.
In section III, we obtain, by perturbation theory, the low energy states of graphene with a sinusoidal modulation of the hopping and identify the eight  zero energy Landau levels that appear  in rippled graphene.
Section IV is devoted to the study of the effect of the electron-electron interaction on the electronic properties. We also present self-consistent results obtained from the Hubbard Hamiltonian. 
Finally, the results are summarized in section V.

%Here we....
\section{Geometry and Hamiltonian}
\begin{figure}[htbp]
\includegraphics[width=9.cm,clip]{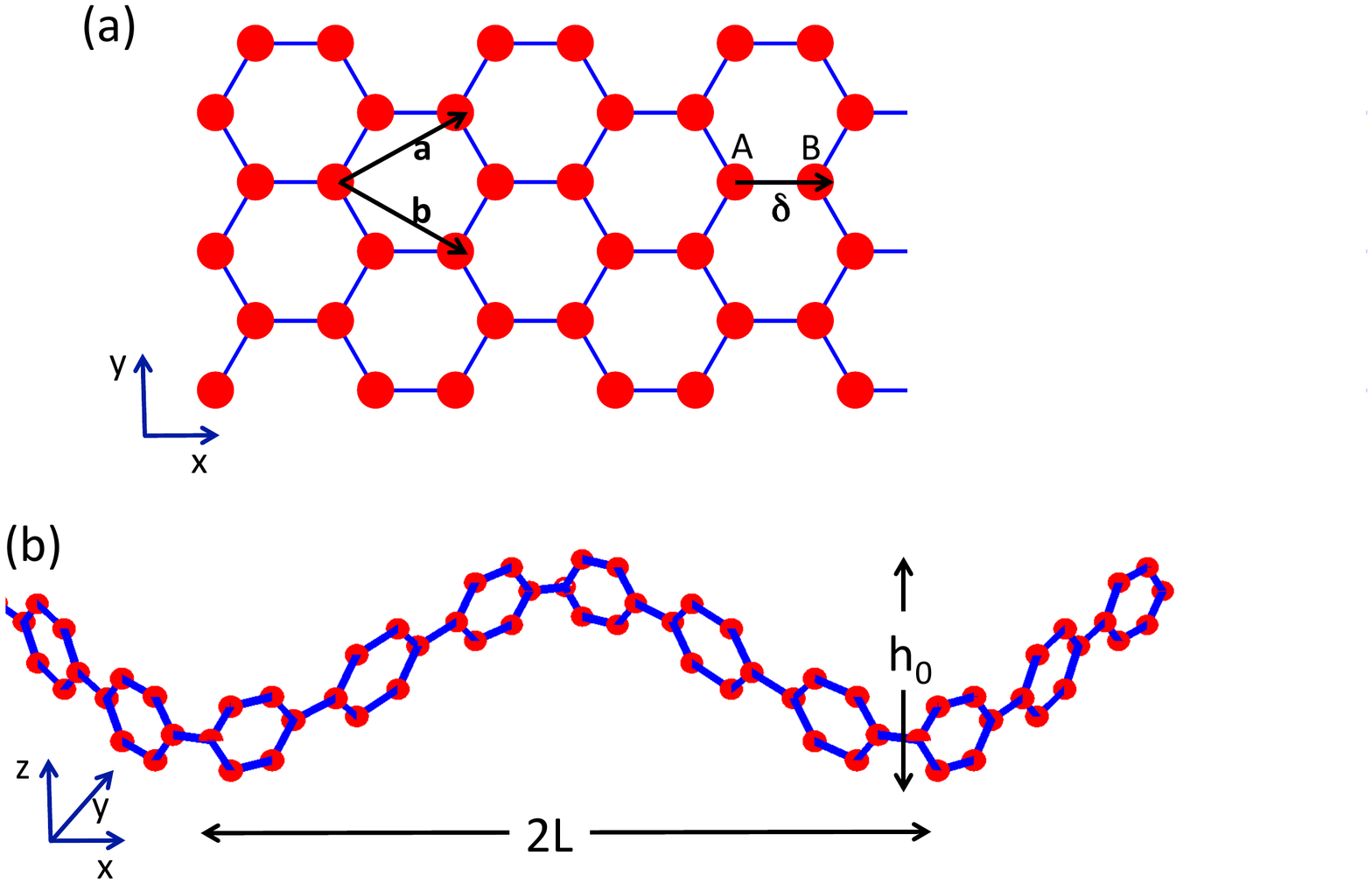}
\caption{(Color online) (a) Graphene crystal structure, ${\bf a}$ and ${\bf b}$ indicated the lattice vectors. ${\boldsymbol \delta}$ is the vector connecting the two atoms, A and B, in the unit cell. (b) Schematic view of a graphene ripple of period 2$L$ and height $h_0$.}                                     
\label{Scheme}
\end{figure}

{\it Geometry.}
In graphene the carbon atoms crystallize  in a two-dimensional triangular lattice with a basis constituted  by two equivalent atoms $A$ and $B$. 
The lattice is defined, see Fig.\ref{Scheme}, by the vectors ${\bf a} =\frac a 2 (\sqrt{3},1)$ and ${\bf b} =\frac a 2 (\sqrt{3},-1)$, and the atoms of the basis are located at the origen, $(0,0)$ and at 
${\bf \delta} =( a /\sqrt{3},0)$, here $a=2.46$\AA  is the lattice parameter. As discussed in the introduction, the ultrathin nature of graphene makes it flexible against out-of-plane deformations of the lattice. Here, we consider a one-dimensional periodic graphene ripple, which modulates the height (z-coordinate) of the carbon atoms according to the expression 
\begin{equation}
h(x)=h_0 \sin{\frac {\pi}L x} 
\label{altura}
\end{equation}
where ${\bf r}=(x,y)$ is the position of the carbon atoms, $h_0$ is the height amplitude and the period is $L$/2, as schematically shown in  Fig.\ref{Scheme}.

{\it Tight-Binding Hamiltonian.}
The electronic properties of graphene are well described by a nearest neighbor tight-binding Hamiltonian\cite{Guinea_2009,Katsnelson-book} of the form
%\begin{equation}
%H_0 = -\sum_{<i,j> ,\sigma } t_{i,j} c^+ _{i,\sigma} c_{j,\sigma}, ,
%\label{TB}
%\end{equation}
%where $c^ + _{i,\sigma}$ creates an electron at site $i$ with spin $\sigma$. 
\begin{equation}
H_0 = -\sum_{<i,j> ,\sigma } \left (t_{i,j}  c^+ _{i,\sigma,A} c_{j,\sigma,B},+h.c \right) ,
\label{TB}
\end{equation}
where $c^ + _{i,\sigma,\alpha}$ creates an electron at lattice site $i$ with spin $\sigma$ and sublattice $\alpha =A, B$.

In pristine graphene the distance between first neighbor carbon atoms is the same along the entire crystal and the hopping  between atoms is constant  $t_{i,j}=t_0$, with $t_0=2.8eV$.
In rippled graphene the corrugation modifies the distance between carbon atoms. When the period of the ripple is much larger than $a$, the distance between two first neighbor carbon atoms $i$ and $j$ located at ${\bf r}_i$ and ${\bf r }_j$ is modified to
\begin{equation}
d_{i,j}\approx d_0( 1 +\frac {\pi ^2}  2 \frac {h_0^2}{L^2} \frac {(x_j-x_i)^2}{d_0 ^2}  \cos ^2 {\frac {\pi} L x_i})
\label{dij}
\end{equation}
being $d_0=a / \sqrt{3}$ the equilibrium distance between carbon atoms. 

The hopping amplitude between nearest neighbor atoms scales with a power law with the length of the bond between atomic centers\cite{Harrison_book,Brey_1983,Brey_1984},
\begin{equation}
t_{i,j} = t_0 \left ( \frac {d_0}{d_{i,j}} \right ) ^{\beta}\approx t_0 \left ( 1 - \frac {\beta \pi ^2 } 2 \frac {h_0^2}{L^2} \frac {(x_j-x_i)^2}{d_0 ^2}  \cos ^2 {\frac {\pi} L x_i} \right ) 
\label{tij}
\end{equation}
being $\beta \sim 2-3$. Therefore, 
%apart from a global reduction of the hopping amplitude, 
a ripple with period $L/2$ induces a modulation in the hopping amplitude with period $L$. The modulation of the hopping is proportional to  $ \sim 10 \frac {h_0^2}{L^2}$, that for some experimental systems can be as larger as 0.1\cite{Meng_2013}.  
%For the shake of clarity, 
%in the rest of the paper we neglect the constant change of the tunneling  and we analyze graphene in presence of a tunneling modulation of the form, 
%\begin{equation}
%t_{i,j} =t_0 \left ( 1 + \delta _t  \frac {(x_j-x_i)^2}{d_0 ^2}  \sin{ G x_i} \right ) 
%\label{tij}
%\end{equation}
%with $G=2\pi/L$ and $\delta _t=\frac {\beta \pi ^2 } 2 \frac {h_0^2}{L^2} t _0$.

{\it Dirac Hamiltonian.}
The low energy, long distance electronic excitations of graphene occur near the celebrated Dirac points, ${\bf K }$=$\frac {2 \pi} a(\frac 1 {\sqrt{3}},\frac 1 3)$ and
${\bf K }'$=$\frac {2 \pi} a(\frac 1 {\sqrt{3}},-\frac 1 3)$. Near these points,  the band structure of graphene is very well described  by Dirac-like Hamiltonians,
 \begin{equation}
 H_D = \hbar v_F  \left ( i   s \partial _y \tau _x - i \tau _y \partial _x \right )  
% \begin{array} {cc}
% 0   & -ik_x -sk_y \\
% ik_x -s k_y & 0 
% \end{array}
% \right ) 
% \label{HD}
 \end{equation}
where  $v_F= \frac {\sqrt{3}} 2  t_0 a $,  and  ${\boldsymbol \tau}$=$(\tau_x,\tau_y,\tau_z)$ are the Pauli matrices acting on a spinor that define the amplitude of the wave function on the sublattices $A$ and $B$ of graphene   and $s$=$+1$ and $s$=$-1$  indicates the Dirac cone ${\bf K}$ and ${\bf K}'$ respectively. In presence of a hopping modulation of  the form of equation \ref{tij}, the modified Dirac Hamiltonians take the form \cite{Guinea_2008},
\begin{equation}
H_D =  \hbar \bar{v}_F (x) \left (i   s \partial _y \tau _x - i \tau _y \partial _x \right )  
+(t_{\parallel}(x)-t_{\perp} (x)) \tau _x
\end{equation}
where $t_{\parallel}(x)$ and $t_{\perp} (x)$ are the hopping parameters  corresponding to the horizontal and oblique bonds, see Fig.\ref{Scheme} and
$\bar{v}_F (x) = \frac {\sqrt{3}} 2  a t _{\perp} (x)$.  We asume that the 
period of the ripple is  much larger than the graphene lattice parameter and  electronic states coming from different Dirac cones do not mix. 
 The modulation of the hopping has two effects on the electronic properties of graphene, it modulates spatially  the Fermi velocity and  creates a position dependent gauge magnetic field\cite{Guinea_2008,Vozmediano_2010,Juan_2012,Juan_2013,Amorim_2016}.

{\it Magnetic gauge field.}
In this work we are interested on the effect of the gauge field on the graphene electronic properties, therefore in order to avoid effects related with the modulation of the Fermi velocity, we only consider modification in the horizontal hoping $t_{\parallel}$ and we take $t_{\perp}$=$t_0$ and the Dirac Hamiltonian takes the form\cite{Shift}
\begin{equation}
H_D ({\bf r})  =
 \hbar v_F \left (i   s \partial _y \tau _x - i \tau _y \partial _x \right )  
+\delta _t \sin{ G x}\, \tau _x
\label{HD_ripple}
\end{equation}
with $G=2\pi/L$ and $\delta _t=\frac {\beta \pi ^2 } 4 \frac {h_0^2}{L^2} t _0$.

With this assumption the modulation of the hopping results in the appearance of the vector potential $A_y$=$s \frac {c \delta _t }{e v_F}\sin {Gx}$. The vector potential has opposite sign in different Dirac points, so that  time reversal symmetry is preserved.  For a given Dirac cone, the effective vector potential  oscillates in space forming alternating regions of positive and negative pseudo magnetic fields, $B_z$=$s\frac {c \delta _t }{e v_F} G \cos {Gx}$.  The magnetic length corresponding to the maximum of the pseudo magnetic field is $\ell=\sqrt{\frac {\hbar v_F}{\delta _t G }}$.
A wave function in the $n=0$ pseudo Landau level should be localized in the region where the pseudo magnetic field has a defined sign, i.e. $\ell < L/2$. That implies that, for observing physical effects related with the pseudo Landau levels quantization,  the parameters describing the  ripple should satisfy,
\begin{equation}
\frac{ 2 \hbar v_F }{\pi L \delta_t} < 1 \, \, \, \, {\rm or} \, \, \, {\rm equivalently} \, \, \,  \frac {4 \sqrt{3}}{\beta \pi ^3} \frac {L a}{h_0^2} <1 \, \, \, \, \, .
\end{equation}
For values $\delta_t \sim 0.05t_0 -0.1t_0$ and a period of the hopping modulation $L =200a$, the effective magnetic length takes values in the 
interval 40$a$-56$a$, smaller than $L/2$,  that correspond to  effective magnetic fields in the range of 27T-38T.

{\it Symmetry Considerations.} We are interested in obtaining the energy spectrum of the system as function of the momentum ${\bf k}$,
that in the following we define with respect to the Dirac points.
%$k_x$ and $k_y$ the momentum measured with respect the Dirac points
We begin by considering some symmetries of graphene in presence of a ripple. We have already mentioned that because the gauge magnetic fields have opposite sign on opposite Dirac points,  time reversal symmetry is preserved.  On the other hand the Hamiltonian does not depend on the coordinate $y$, and therefore the momentum
$k_y$ is a good quantum number and the eigenfunction can be written as $\varphi _{k_y} ({\bf r})= \psi _{k_y} (x) e ^{i k_y y }$, where $\psi_{k_y} (x)$ is an eigenfunction of
the effective Hamiltonian $H_D (x,k_y )=e ^{-i k_y y} H_D({\bf r}) e ^{ik_y y}$. Besides, the  low energy Dirac Hamiltonian, Eq.\ref{HD_ripple} satisfies the relation $\tau _z H_D (x)\tau _z=-H_D(x)$, and this implies that for any eigenstate with momentum  
${\bf k}$  and energy $E({\bf k})$ there is a state with opposite energy and the same momentum. Finally, the Hamiltonian has the property
$H_D (-k_y,x+ L/2)= \tau_z H^*_D (k_y, x) \tau_z$ that implies that for any zero energy state appearing  at a particular $k_y$, there exits another zero mode at momentum $-k_y$.

\section{Perturbation theory.}
%The low-energy physics of strained graphene is well described by the Dirac Hamiltonian given by equation \ref{HD_ripple} which includes the periodic modulation of the hoppings here considered. 
%In the vicinity of the Dirac points, due to presence of the ripple induce psudomagnetic fields, the behavior of the electrons is modified, giving rise to a Landau level-like quantization, resulting
%in the so-called pseudo Landau levels experimentally observed \cite{Levy_2010} These low-energy pseudo Landau levels are   highly degenerate and are prone to interaction instabilitiesi \cite{ Venderbos_2016}. The behavior of these states is important  to understand different aspects of graphene properties. We make an analysis of the wave fuctions corresponding to these low energy states close to the Dirac points, for different values of the ripple amplitude and period. The agreement  with results obtained from tight-binding calculations allows to confirm the validity of the  perturbation solutions for wavevectors close to $-{\bf K}$and $+{\bf K}$.      

The low energy Hamiltonian, Eq.\ref{HD_ripple}, when $k_y$=$0$  presents  zero energy states with the explicit form, $\psi ^1 _{k_y=0}=(e^{\bar{\delta_t} \cos{Gx}},0) ^\dagger$ and
 $\psi ^2_{k_y=0}=(0,e^{-\bar{\delta_t} \cos{Gx}}) ^\dagger$ that for small values of the magnetic length $\ell$ take the form of gaussians  centered at $x$=$nL$ and $x$=$L/2+nL$ respectively and have amplitude just in one of the graphene sublattices, here $n$ is an integer and
 $\bar{\delta_t}$=$\frac {\delta _t  }{\hbar v_F G}$. For finite values of $k_y$ it is not possible to obtain analytical solutions, but  we expect  the wavefunctions $\psi ^1_{k _y}$ and $\psi ^2 _{k_y}$ to be centered at
$x$=$s k_y  \ell ^2 $ and $x$=$L/2$-$ s k_y \ell ^2$ respectively. Therefore for small values of $k_y$  we choose the  basis 
\begin{eqnarray}
\psi ^1 _{k_y} & = & \frac 1  { \sqrt{I_0 (2 \bar{\delta _t}) } }\left ( \begin{array}{c} e^{\bar{\delta} \cos{G(x-sk_y \ell ^2)}} \\ 0 \end{array} \right )  \nonumber \\ \psi ^2 _{k_y} & = & \frac 1  { \sqrt{I_0 (2 \bar{\delta _t}) } }\left ( \begin{array}{c} 0 \\ e^{-\bar{\delta} \cos{G(x+sk_y \ell ^2)}}  \end{array} \right )
\label{basis}
\end{eqnarray}
where $I_0$ is the modified Bessel function of the first kind of zero-order.
For small wave vector the wavefunctions $\psi _1$ and $\psi _2$ are similar to the  zero energy real magnetic field Landau levels \cite{Goerbig_2011}. However the structure of the spinors is different.
For the valley ${\bf K}$ the wavefunction  $\psi ^1 _{k_y}$ has only amplitude in sublattice $A$ and  it is centered at positions near $x=nL$,  where the pseudo magnetic field is positive, $B_{ef} ^+$, on the contrary $\psi ^2 _{k_y}$ has only support in sublattice $B$ and is located in the regions near $x$=$nL+L/2$ where  the   effective magnetic field is negative, $B_{ef}^-$. For the Dirac cone ${\bf K}'$, the spatial locations  of the wavefunctions $\psi ^1 _{k_y}$ and $\psi ^2 _{k_y}$ are reversed with respect the ${\bf K}$ valley, so that time reversal symmetry is preserved.

The  Hamiltonian of Eq.\ref{HD_ripple} projected in the basis given by Eq.\ref{basis} takes the form, 
\begin{equation}
\bar H = \left (
\begin{array}{cc}
0 & t(k_y)\\
t(k_y)  & 0
\end{array}
\right )
\label{Heff}
\end{equation}
with 
\begin{equation}
t(k_y)= \frac{-\hbar v_F k_y  { \bar I} _0+ \delta _t { \bar I}_1 \left (1-\cos(Gk _y \ell ^2) \right ) {\rm sgn}(k_y) } {I_0 (2 \bar{\delta _t})}
\label{teff}
\end{equation}
where  ${\bar I} _n = I_n(\bar {\delta _t} \sin (G k_y \ell^2))$ being $I_n$ the modified Bessel function of the first kind of order $n$.  
At small momenta $k_y$ the dispersion is lineal with renormalized velocity
\begin{equation}
\bar  v_F = v_F / I_0(2 {\bar \delta _t}) \, \, ,
\label{vfrenor}
\end{equation}
 which, although finite, decreases exponentially when increasing  $L$ or $\delta _t$. 
As a result of that, the  band structure obtained in numerical calculations \cite{Guinea_2008,Wehling_2008} shows apparent dispersionless degenerate   pseudo Landau levels. 
For larger values of $k_y$  the
overlap between the wavefunctions $\psi ^1 _{k_y}$ and $\psi ^2 _{k_y}$ increases and the pseudo Landau levels acquire a dispersion. 
Therefore, at zero energy, there are eight almost degenerated pseudo Landau levels, Fig.\ref{order}, denoted by
$|A,B_{ef}^+,\sigma>$, $|B,B_{ef}^-,\sigma>$, $|A,B_{ef}^+,\sigma>'$ and $|B,B_{ef}^-,\sigma>'$, in such a way that in regions with $B_{ef} ^{+(-)}$ the states have  amplitude only in sublattice $A(B)$. 
Here $\sigma$ is the projection of the electron spin.

\begin{figure}[htbp]
\includegraphics[width=8.5cm,clip]{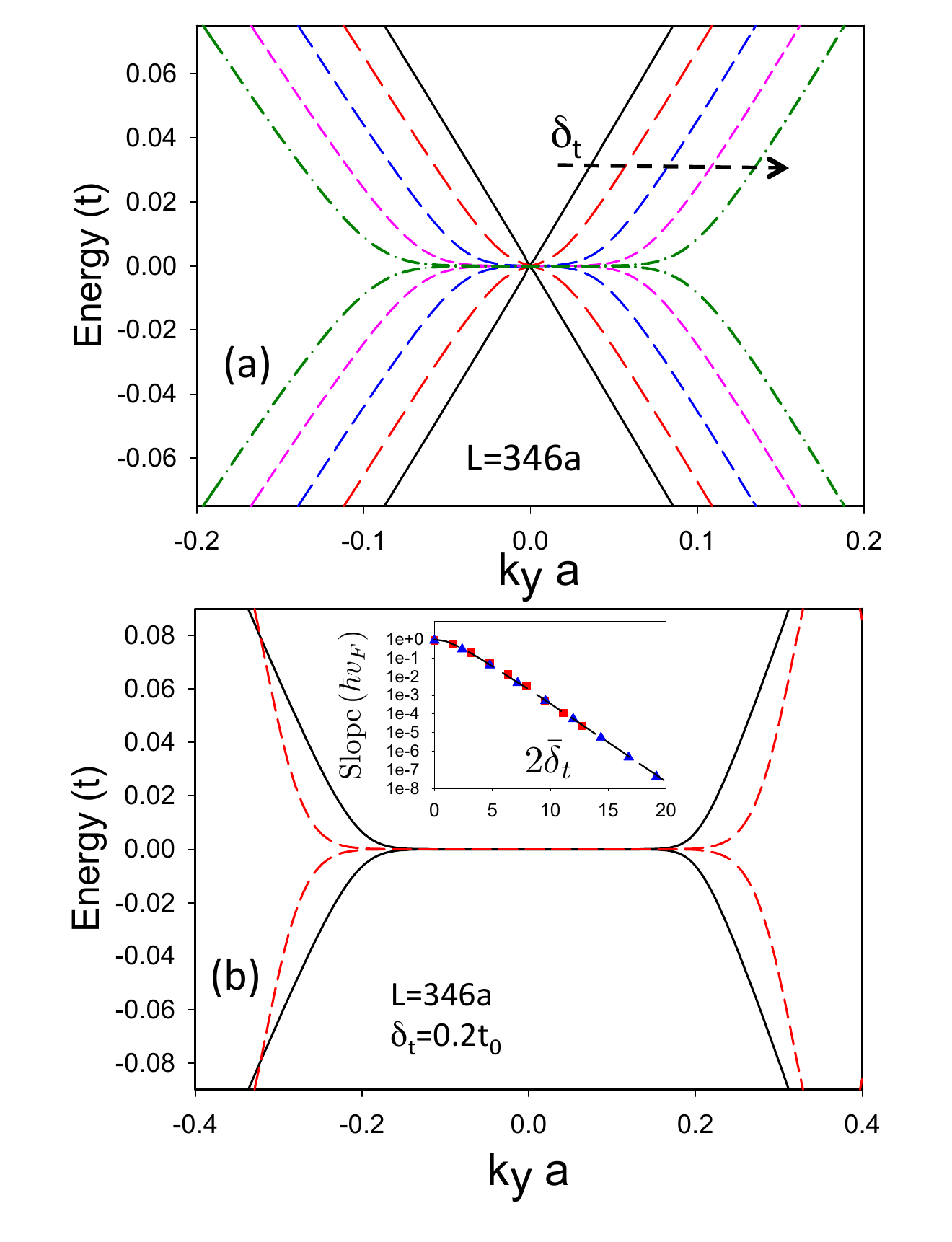}
\caption{(Color online) (a)Band structure near a Dirac point for graphene in presence of  a sinusoidal modulation of the hoping with period $L$=246$a$ and $\delta_t$ running from 0 to 0.1$t$.
% (b)Band structure of hopping modulated graphene
(b) same as (a) for $L=346a$ and $\delta _t$=0.2$t$. Continuous lines correspond to tight binding results whereas dashed lines are the dispersion obtained using Eq.\ref{Heff}.
Inset in (b) shows, as function of the dimensionless parameter ${\bar \delta} _t$,  the renormalized  Fermi velocity at Dirac points for $L=178a$ (triangles)  and $L=115a$ (squares) and values of $\delta _t$ ranging from zero to $\delta _t=0.2t$. The dashed line corresponds to expression Eq.\ref{vfrenor}. }                                     
\label{Slope}
\end{figure}
\begin{figure}[htbp]
\includegraphics[width=8.cm,clip]{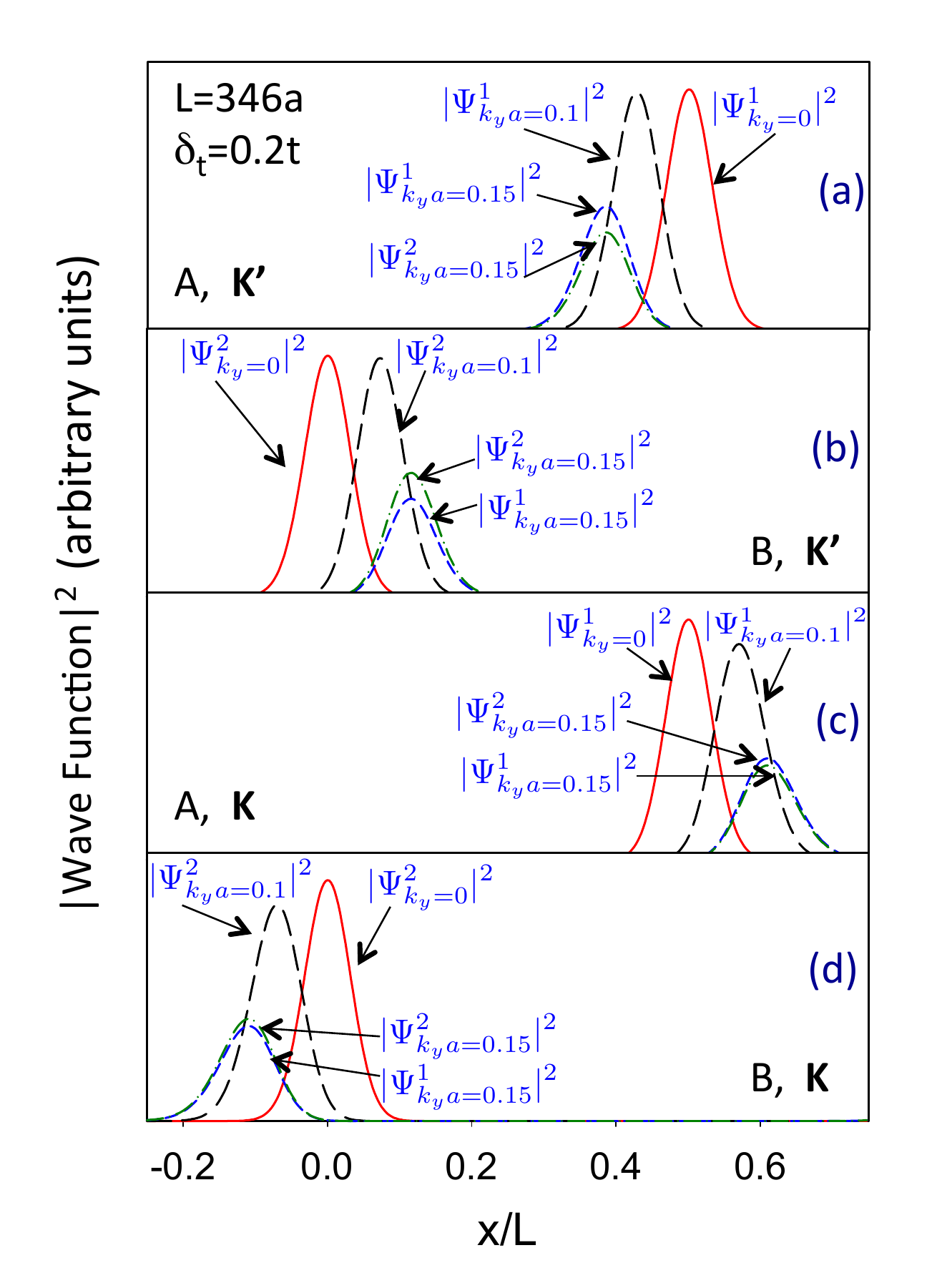}
\caption{(Color online) Square of the wavefunctions obtained from the tight binding calculation for wavevectors near the ${\bf K'}$(a-b) and ${\bf K}$(c-d) points. The momentum $k_y$ is measured with respect the Dirac points.   For each $k_y$ we consider the two states, $1$ and $2$ closest to zero energy. Panels (a) and (c) correspond to the amplitude on sublattice $A$ whereas panels (b) and(d) correspond to amplitude on sublattice $B$. The results are obtained in the tight-binding approximation with a sinusoidal modulation of the hopping of period $L$=346$a$ and amplitude $\delta _t$=0.2$t$.
}                                  
\label{wfU0}
\end{figure}

{\it Tight Binding results.}
In order to check the results obtained with the modified Dirac equations, we have performed microscopic tight binding calculations.  We have considered a sinusoidal modulation of the hopping 
in a supercell along the $x$-direction with the geometry presented in Fig.\ref{Scheme}.  In this geometry the Brillouin zone of the supercell is rectangular; in the $y$-direction goes form $0$ to $ {\frac {2 \pi} a}$ and in the $x$ direction from $0$ to $\frac {2 \pi} L$. For large values of $L$, the eigenvalues do not depend on $k_x$ and we just consider $k_x=0$. 
In this geometry the original Dirac points ${\bf K}'$ and ${\bf K}$ are folded at the wavevectors $(\frac {2 \pi} {3a},0)$ and $(\frac{4 \pi} {3a},0)$ respectively.
As pointed out above, in order to avoid effects related with the spatial dependence of the Dirac velocity, we  only modulate the hopping in the horizontal bonds. In Fig.\ref{Slope}(a) we plot the lowest energy conduction band and the highest energy valence band of rippled graphene for a period modulation $L=346a$ and different values of the amplitude modulation $\delta _t$. As predicted by the Dirac equation the linear dispersion  becomes  flatter as $\delta _t$ increases and, for moderate values of the modulation, the dispersion reminds that of Landau levels. In the inset of Fig.\ref{Slope}(b) we plot the slope of the dispersion at $k_y \rightarrow 0$ as function of the dimensionless parameter $2\bar \delta _t$, for two different values of the period, $L=178a$ and $L=115a$ and different values of $\delta _t$. In agreement with the continuous solution, we obtain that the velocity only depends on the the combination $\delta _t L$ and is determined with a high precision by 
the renormalized velocity given in Eq.\ref{vfrenor}. In Fig.\ref{Slope}(b) we compare  the band structure  for a hopping modulation of period $L$=346$a$ and amplitude $\delta_t$=$0.2t$, as obtained from the tight-binding approximation and from the  approximated  solution of the  Dirac equation,  Eq.\ref{Heff}. 
The perturbation solution describes qualitatively the almost dispersionless states that occur near the Dirac points for small wavevectors, where the overlap between the wavefunctions from Eq.\ref{basis} is practically null and how, for large enough values of $k_y$, the basis wave functions overlap and the coupling between states induces a bonding-antibonding splitting. The trial wavefunctions given in Eq.\ref{basis},  for small wave vectors,  qualitatively describe the solutions of the tight-binding Hamiltonian. In Fig.\ref{wfU0} (a)-(b) we plot, for wavevectors near ${\bf K}'$ the square of the wavefunction  on sublattice $A$ and $B$ respectively.  For small values of $k_y$ the almost degenerated gaussian-like wave functions  are located at $x$=$sk_y \ell ^2$ and $x$=$L/2-sk_y \ell ^2$ and have amplitude only in sublattices $A$ and $B$ respectively. As the wave vector $k_y$ increases the coupling between the wavefunctions increases, the energes of the states split and the states  get amplitude on both sublattices. Fig.\ref{wfU0} (c)-(d) represent the same as (a)-(b) respectively, for wavevectors near ${\bf K}$.

\section{Electron-electron interaction.} 
In this section we address the effects  of electron-electron interactions on the zero energy pseudo Landau levels formed in graphene by the modulation of the hopping terms. 
%We will consider the Hubbard model and carry out  numerical calculations investigating the different self consistent solutions obtained for different values of the Hubbard interaction and
%ripples in order to get insight into the different ordered states.
 For large values of the product $L \delta _t$, the high density of states at the Fermi energy implies instabilities of the system again broken symmetry states that open gaps at the Fermi energy.
In the Dirac approximation of graphene, the valley and spin variables are equivalent isospin indices and  ignoring small symmetry breaking terms  due to lattice effects\cite{Alicea_2006,Herbut_2007,Gusynin_2006} and  neglecting the  difference between the inter and intra valley  electro-electron interaction, the Hamiltonian  is SU(4) invariant.  Therefore, in the framework of the quantum Hall ferromagnetism\cite{Moon_1995,Fertig_1994,Fertig_1997},
we expect  the ground 
state to break  spontaneously the SU(4) symmetry putting many electrons into the same pseudospin state and minimizing their exchange energy to the lowest value satisfying the Pauli exclusion principle\cite{Fertig_2006,Yang_2006,Abanin_2006,Goerbig_2006,Yang_2007, Das-Sarma_2009,Jung_2009,Barlas_2012} and  
opening an energy gap in the charge excitations.

These  quantum Hall pseudoferromagnetic ground states are degenerated and broken symmetry terms, lattice effects or Landau level mixing can lift the degeneracy favoring  some particular isospin order.  
In the presence of a real magnetic field, 
mean field calculations using a tight-binding Hamiltonian with a Hubbard term 
obtain a real spin  antiferromagnetic  ground state\cite{Jung_2009} and,   calculations using the
Dirac equation including isospin anisotropy\cite{Aleiner_2007} predict as well an antiferromagnetic order\cite{Kharitonov_2012}.

We study here the electronic and magnetic properties of rippled graphene by obtaining  self-consistently the  tight-binding Hamiltonian describing a sinusoidal modulation of the hopping in the $x$-direction (see Fig.\ref{Scheme}) with  an on-site  Hubbard interaction term of the form,
\begin{equation}
H_U = U \sum _ {i,\alpha} n_{i,\uparrow,\alpha}  \, n_{i,\downarrow,\alpha}
\end{equation}
where $n_{i,\sigma,\alpha}$=$c^+_{i,\sigma,\alpha} c_{i,\sigma,\alpha}$ is the fermionic  number operator for lattice site $i$, spin projection $\sigma = \uparrow, \downarrow$ and sublattice $\alpha$. The usual mean field decomposition of the on-site interaction leads to an effective one-particle interaction term, 
\begin{equation}
H_U \! =\! U \sum _{i,\alpha}\left ( \langle n_{i,\uparrow,\alpha} \rangle n_{i,\downarrow,\alpha}+ \langle n_{i,\downarrow,\alpha} \rangle  n_{i,\uparrow,\alpha}-\langle n_{i,\uparrow,\alpha}\rangle \langle n_{i,\downarrow,\alpha}\rangle  \right ) 
\end{equation}
 $\langle n_{i,\sigma,\alpha} \rangle$ denotes the average occupation operators, that are obtained by solving self consistently the tight-binding Hamiltonian and the mean field Hubbard term. The nature of the self-consistent ground state is characterized by the following order parameters,
\begin{eqnarray}
M_i & = & \langle n_{i,\uparrow,A} \rangle + \langle n_{i,\uparrow,B} \rangle -  \langle n_{i,\downarrow,A} \rangle - \langle n_{i,\downarrow,B} \rangle
\nonumber \\
m_i & = & \langle n_{i,\uparrow,A} \rangle - \langle n_{i,\uparrow,B} \rangle -  \langle n_{i,\downarrow,A} \rangle + \langle n_{i,\downarrow,B} 
\rangle  \, ,
\label{orderp}
\end{eqnarray}
which indicate the local ferromagnetic  and antiferromagnetic order respectively.

\begin{figure}[htbp]
\includegraphics[width=8.cm,clip]{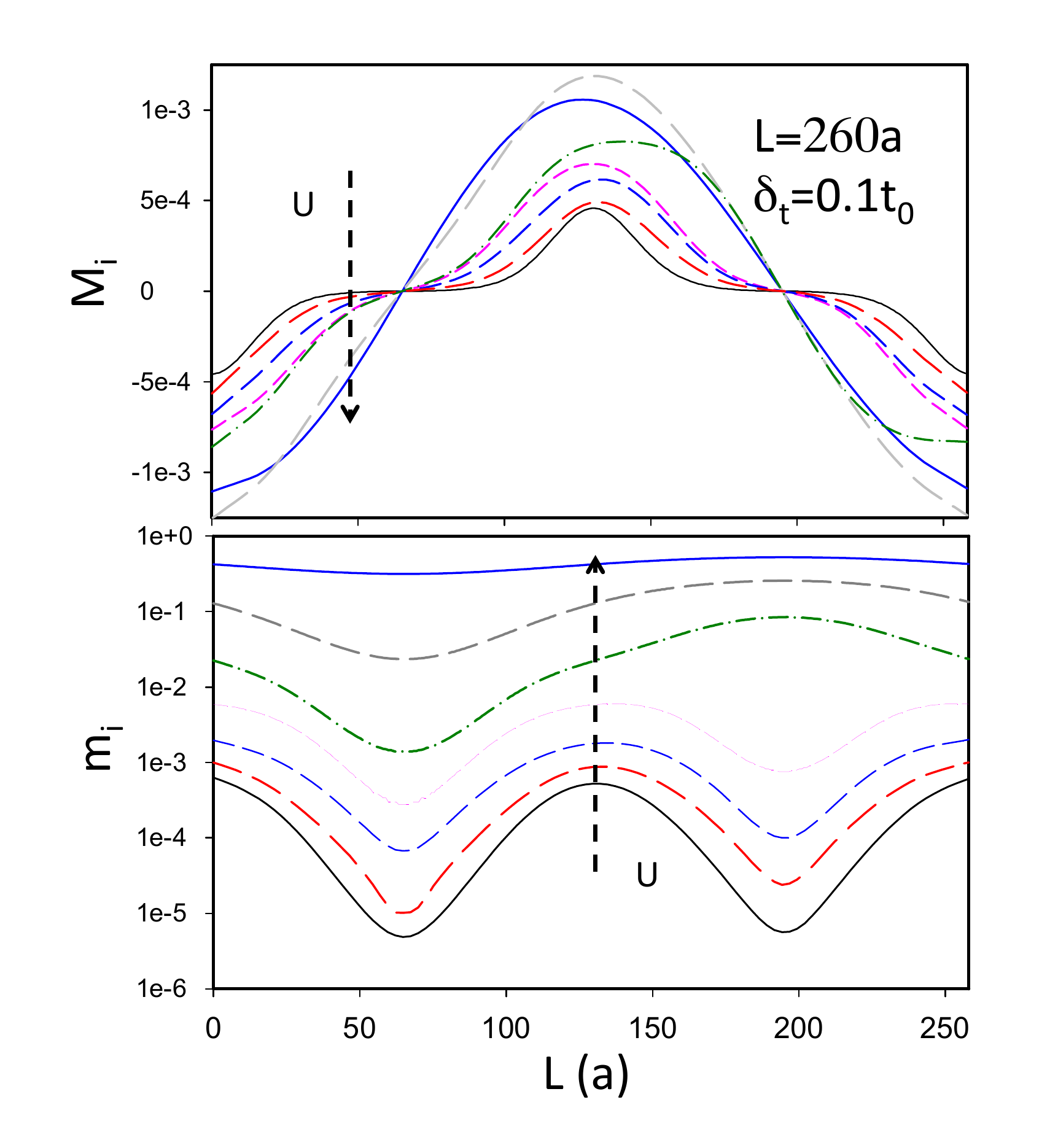}
\caption{(Color online) Ferromagnetic (top) and antiferromagnetic (bottom) local order parameter for a ripple with $\delta _t$=0.1$t_0$, $L$=260$a$ and $U$=0.5$t_0$, $t_0$, 1.5$t_0$, 2$t_0$, 2.2$t_0$, 2.3$t_0$ and 2.5$t_0$. }                                  
\label{Polarization}
\end{figure}

\begin{figure}[htbp]
\includegraphics[width=8.cm,clip]{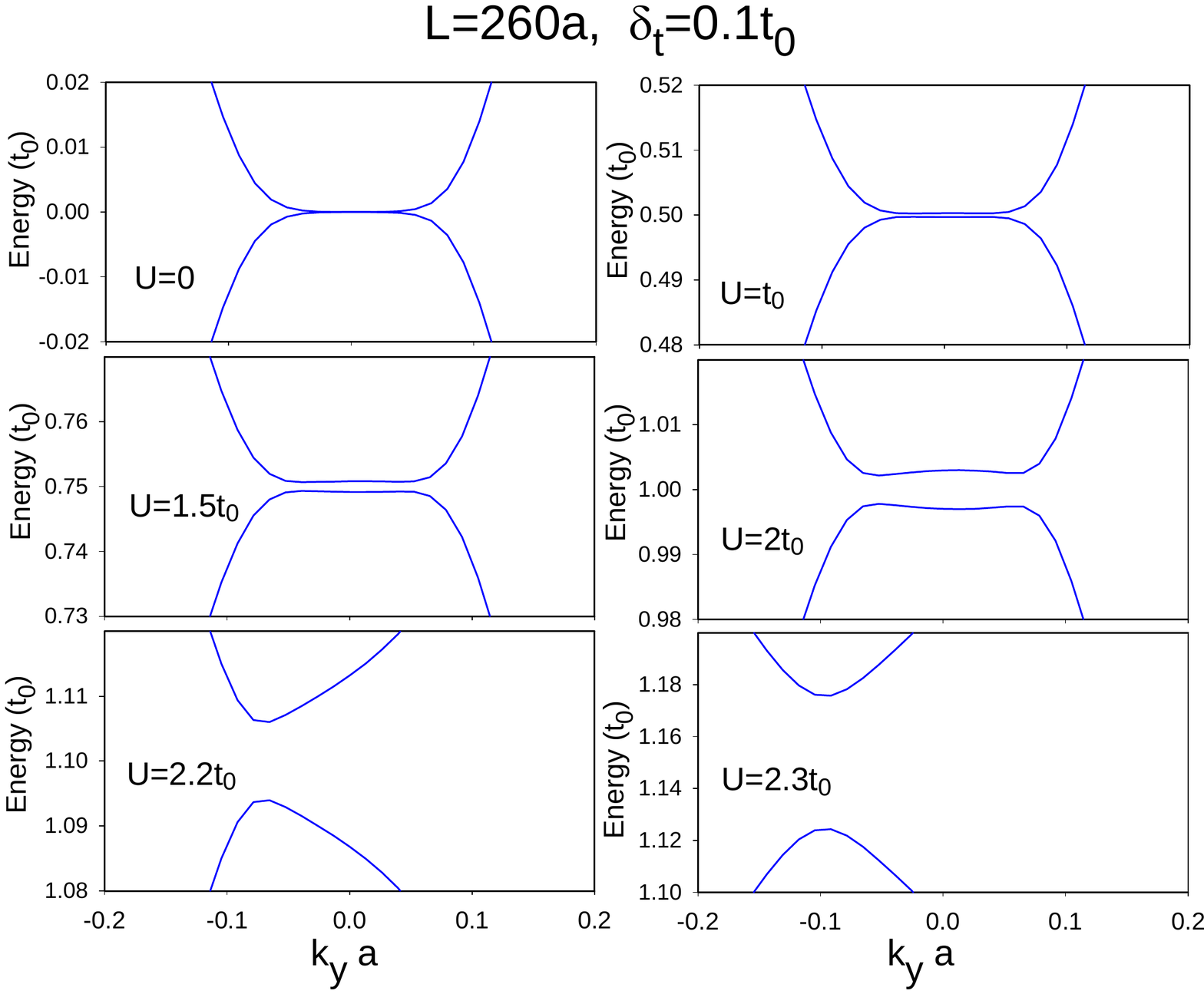}
\caption{(Color online) Band structure near  Dirac point for graphene in presence of a  sinusoidal modulation of the hoping 
with period $L$=246$a$,  $\delta_t$=0.1$t_0$ and different values of the Hubbard coupling. The bands are spin degenerated.} 
\label{BandasFig-600-dr10-dirac}
\end{figure}

We have performed self-consistent calculations for different values  of $U$, $L$, and $\delta _t$. In all the cases we find that the charge is uniformly distributed, i.e. $\langle n_{i,\uparrow,A} \rangle + \langle n_{i,\downarrow,A} \rangle$=$\langle n_{i,\uparrow,B} \rangle + \langle n_{i,\downarrow,B} \rangle$=1.
For a given ripple characterized by a period $L$ and a hopping modulation $\delta _t$, there is a critical value of the Hubbard interaction  $U_C$ for which the system undergoes a second order phase transition to a phase with both local FM and AFM order. In the inset of Fig.\ref{Phase-Diagram} we plot the value of $U_C$ for different values of $L$ and $\delta _t$. Interestingly, all the values of $U_C$ seem to collapse in a unique curve, see  Fig.\ref{Phase-Diagram}, when plotted with respect the dimensionless parameter $\bar \delta _t$ =$\frac {\delta _t}{\hbar v_F G}$.
For small $\bar \delta _t$,  the critical Hubbard parameter practically coincides with the value $U_C^{AF}$=2.23$t_0$\cite{Sorella_1992,Martelo_1996} for which pristine graphene undergoes an antiferromagnetic transition. Note however that, in the case of rippled graphene,  both FM and AFM local order parameters become finite at the phase transition.  For larger values of $\bar \delta _t$ 
the pseudo Landau levels become better defined, the density of states at the Fermi energy increases  and  therefore $U_C$ drops  and eventually, for large enough $\bar \delta _t$,  becomes zero.

In our numerical  calculations we obtain that, for any value of the parameters $\delta_t$, $L$, and $U$,
the total magnetization of the rippled graphene is zero, in addition and because  graphene is  a triangular bipartite lattice, the two sublattices have opposite magnetic polarization.

  Fig.\ref{Polarization}  and Fig.\ref{BandasFig-600-dr10-dirac} show  the magnetic  local order parameters and band structure  respectively,  for a ripple with $\delta _t $=0.1$t_0$, $L$=260$a$ and different values of the Hubbard interaction. For this ripple $U_C \sim 0.1t_0$ so that for the values of $U$ plotted in Fig.\ref{Polarization}  there is always magnetic order. 
The local FM order appears in the   spatial regions $x \sim nL$ and $x\sim n/2+nL$,
where the pseudo effective magnetic fields are larger and the pseudo Landau levels are well defined.  The orientation of the 
spin  polarization in these regions is coupled with  the orientation of the pseudo magnetic  field in such a way that the global  FM order is zero. 

This local FM order results in a splitting of the originally degenerated Landau levels and the opening of an energy gap at the Fermi energy, as can be observed in Fig.\ref{BandasFig-600-dr10-dirac}. The numerical results indicate that  the  originally degenerated eight pseudo Landau levels split into two sets of four degenerated pseudo Landau levels, being the occupied states the centered in the region with positive effective magnetic field and spin down that correspond to wavefunctions with amplitude in sublattice A and the centered in the region with negative effective magnetic field and spin up that correspond to wavefunctions with amplitude in sublattice B as schematically shown in Fig. \ref{order} (a).

Superposed to the FM modulation there is an AFM order that for small values of $U$, is much smaller than the FM order, but that increases with $U$ and for values near $U_C ^{AF}$ dominates over the FM order. 
For these values of $U$, the self-consistent one-particle Hubbard interaction term in the Hamiltonian is much stronger than the term corresponding to the gauge magnetic fields, and the pseudo Landau levels near zero energy are completely washed out in the band structure, Fig.\ref{BandasFig-600-dr10-dirac}.
For these  large values of  $U$, the relevant parameter is the ratio between $U$ and the tunneling, so that  a spatial modulation of  the hopping  reflects in a modulation of the AFM order. As it is shown in Fig.\ref{Polarization}, for values of $U \geq U_C ^{AF}$
the AFM order parameter is larger in the regions with smaller hopping, $x \sim 3 \frac L 4 + nL$ and smaller in the regions
with larger hopping $x \sim \frac  L 4 + nL$.  

\section{Summary}
The two-dimensional geometry of graphene makes it unstable against  buckling and rippling. In this work we have studied the electric and magnetic properties of rippled graphene. Long wavelength ripples induce a modulation of the hopping parameters in the Hamiltonian  of graphene, that translates in the appearance of gauge magnetic fields and pseudo Landau levels. By applying  perturbation theory to the modified Dirac equation  we have characterized 
the eight, spin and valley degenerated,  zero energy  pseudo Landau levels that appears  in rippled graphene. 
For both Dirac cones and  spin orientation wavefunction in pseudo Landau levels in regions with positive gauge magnetic field have amplitude only in 
sublattice $A$ whereas in regions with negative field have weight only on the opposite sublattice $B$. The high degeneracy at the Fermi energy makes the system prone to interaction instabilities.  We have solved self consistently the Hubbard model applied to the 
tight-binding graphene Hamiltonian and we have found that for moderate values of the Hubbard interaction,  the system becomes a  gapped quantum Hall local  pseudo ferromagnetic state
with  positive spin polarization in regions with positive effective magnetic field and  with the opposite polarization
in regions with negative effective gauge field.  On top of this local FM order there is an antiferromagnetic order, that increases as $U$ increases and   for values of the interaction near $U_C^{AF}$  the system becomes antiferromagnetic with a local  order parameter modulated by the amplitude of the hopping.
This local pseudo ferromagnetic state, is different from the ground state that will appear in presence of a real magnetic field, and in the mean field Hubbard model is a N\'eel state.

%The low-energy physics of strained graphene is well described by the Dirac Hamiltonian given by equation \ref{HD_ripple} which includes the periodic modulation of the hoppings here considered. 
%In the vicinity of the Dirac points, due to presence of the ripple induce psudomagnetic fields, the behavior of the electrons is modified, giving rise to a Landau level-like quantization, resulting
%in the so-called pseudo Landau levels experimentally observed \cite{Levy_2010} These low-energy pseudo Landau levels are   highly degenerate and are prone to interaction instabilitiesi \cite{ Venderbos_2016}. The behavior of these states is important  to understand different aspects of graphene properties. We make an analysis of the wave fuctions corresponding to these low energy states close to the Dirac points, for different values of the ripple amplitude and period. The agreement  with results obtained from tight-binding calculations allows to confirm the validity of the  perturbation solutions for wavevectors close to $-{\bf K}$and $+{\bf K}$.      

%We will consider the Hubbard model and carry out  numerical calculations investigating the different self consistent solutions obtained for different values of the Hubbard interaction and
%ripples in order to get insight into the different ordered states.

 \begin{acknowledgments} L.B acknowledges support from  MINECO/FEDER under grant
FIS2015-64654-P. 
M.P.L.S. acknowledges financial support by the Spanish MINECO grant
FIS2014-57432-P, the European Union structural funds
and the Comunidad de Madrid MAD2D-CM Program (S2013/MIT-3007).

\end{acknowledgments} 

%\bibliography{ref}

\begin{thebibliography}{63}
\bibitem [{Shi()}]{Shift}%
  \BibitemOpen
  \href@noop {} {}\bibinfo {note} {We neglect a constant term proportional to
  $\tau _x$ that can be taken away by a gauge transformation.}\BibitemShut
  {Stop}%
\end{thebibliography}

\end{document}